\documentstyle[11pt,twoside,amsfonts,ntsha_win]{article}
\RequirePackage[OT1,T2A]{fontenc}
\RequirePackage[cp1251]{inputenc}
\RequirePackage[english,ukrainian]{babel}
\addto\captionsukrainian{
}
\title[Варіяційність та приспішення]{ВАРІЯЦІЙНІСТЬ З ПОХІДНИМИ ДРУГОГО ПОРЯДКУ, РЕЛЯТИВІСЬКЕ ПРИСПІШЕННЯ ТА ВЗАЄМОДІЯ
„ДЗИҐИ“
 З ТЕНЗОРОМ КРИВИНИ У ДВО-ВИМІРНОМУ ПРОСТОРІ-ЧАСІ}
\author[Р.~Я.~Мацюк]
{Роман МАЦЮК
}
\date{Інститут прикладних проблем механіки і математики\\ НАН України,\\
 вул.~Наукова, 3$^{\mbox б}$, Львів 79000 \footnotetext{PACS~2006~numbers
11.15.Kc, 02.40.Ky, 45.20.Jj, 45.50.-j} \footnotetext{Стаття подається в авторській редакції}}
\begin{document}
\setcounter{page}{542}
{\renewcommand{\baselinestretch}{1.2}
 \maketitle
}
 {\hfill \small Редакція отримала статтю 15 грудня 2008~р.}
\thispagestyle{myheadings}
\markboth{~\hrulefill~Фізичний збірник НТШ т.7 2008 p.}
     {Фізичний збірник НТШ т.? 200? p.~\hrulefill~}

\begin{abstract}Подаємо варіяційний опис геодезійних кіл у двовимірному ріманівському многовиді. Вивчаємо
пов'язання з рухом одновимірної релятивіської „дзиґи“ і взагалі з релятивіськи--сталоприспішеним рухом.
\end{abstract} {
\newcommand{\msp}{{\kern.03em}}
\newcommand{\ep}[2]{{\epsilon_{#1#2}}}
\newcommand{\om}[2]{{\omega_{#1#2}}}
\newcommand{\ga}[3]{{\Gamma^{#1}{}_{#2#3}}}
\newcommand{\dg}{{\sqrt{|g|}\,}}

\newcommand{\bdt}{^{{\displaystyle\boldsymbol{\cdot}}}}
\newcommand{\SP}[2]{{S'{}_{#1#2}}}
\newcommand{\bu}{{u}}
\newcommand{\bw}{{{u'}}}
\newcommand{\bp}{{\pi}}
\newcommand{\bq}{{\pi^{(1)}}}
\newcommand{\bpr}{{{'}}}
\newcommand{\pr}[2]{{{#1\!\cdot\!#2}}}
\newcommand{\pw}[2]{\left({#1}\wedge{#2}\right)}
\newcommand{\nbu}{{\left\|{u}\right\|^{\mathstrut}}}
\newcommand{\nw}[2]{{\left\|{#1}\wedge{#2}\right\|}}
\newcommand{\bcdot}{{{\cdot}}}
\newcommand{\N}[1]{{\left\|{u}\right\|^{\mathstrut\scriptscriptstyle#1}}}
\newcommand{\NS}[1]{{\left\|{S}\right\|^{\mathstrut\scriptscriptstyle#1}}}
\newcommand{\uu}{{{(u\!\cdot\!u')}}}
\newcommand{\SS}[1]{^{\mathstrut\scriptscriptstyle#1}}
\let\dfrac\frac
\newtheorem{PRP}{Річ}
\newtheorem{CMM}{Об'яснення}
\newtheorem{REM}{Зауваження}
\section{ВСТУП}
\paragraph{\bf Приспішення та гіперболічна нить.}
Геодезійним колом в геометрії зветься сплющена нить всюди сталої першої Френе--кривини~$k$. „Сплющена“ означає, що
уздовж ниті всі повищі крив\'ини є зеро. Дивним чином це поняття збігається, незалежно від вимірів простору, з
фізичним означенням сталоприспішеного руху~\cite{HILL1945}: графіком релятивіського сталоприспішеного руху є
сплющена світова нить сталої кривини~$k$, яка в цьому випадку зветься прискоренням пробної частки. Спільним
визначальним рівнянням, як геодезійного кола, так і світової ниті рівноприспішеного руху, записаним в природному
в\'ідмірі
\begin{equation}\label{u2=1}
g_{mn}\frac{Dx^m}{ds}\frac{Dx^n}{ds}=1\,,
\end{equation}
є таке~\cite{matsyuk:Yano131}:
\begin{equation}\label{1.1}
  \frac{D^3x^l}{ds^3}+g_{mn}\frac{D^2x^m}{ds^2}\frac{D^2x^n}{ds^2}\frac{Dx^l}{ds}=0.
\end{equation}
\paragraph{\bf „Дзиґа.“}
Частина людей вірить, що рух релятивіської дзиґи потрібно описувати рівнянням Матісона (гляди уклад~(3.29) з
письма~\cite{matsyuk:4}, а також п\'исьма \cite{maciuk:Ragusa}, \cite{maciuk:Plyatsko} та~\cite{maciuk:Natario}):
\begin{equation}\label{matsyuk:5}
M\frac{Du^n}{ds} =\frac{D^2u_{m}}{ds^2}\, S^{nm} - {\frac{{1}}{{2}}}R^{n}{}_{mkl} u^{m}S^{kl}{\rm \,.}
\end{equation}
Це рівняння, окрім природного відміру вздовж світової ниті власним часом, передбачає ще й додаткову {\it умову
Матісона--Пірані\/} (гля\-ди ~\cite{matsyuk:4}, с.~182), яка повинна супроводжувати будь--які перетворення, з ним
пов'язані:
\begin{equation}\label{matsyuk:MP}
u_{n} S^{nm} = 0{\rm .}
\end{equation}
У двовимірному світі „інтеґральна“ умова Матісона--Пірані~(\ref{matsyuk:MP}) просто забороняє будь-який розвиток.
Далі у Відділі~\ref{Dixon} спробуємо на варіяційній основі так переробити рівняння Матісона~(\ref{matsyuk:5}),
щоби рівняння рівноприспішеного руху стало його узагальненням.

\section{ВАРІЯЦІЙНЕ РІВНЯННЯ ДЛЯ ГЕОДЕЗІЙНИХ КІЛ}
\paragraph{Позначки.}
Коли записуємо світову нить~$\sigma\subset\frak M$ (в геометричних термінах -- одновимірний підмноговид
(лже)ріманівського простору~$\frak M$) за допомогою функцій $x^{n}({\xi})$, говоримо про {\it стежку}
$\tau\,:\xi\mapsto x^{n}(\xi)$ у многовиді~$\frak M$, образом якої є ця нить, иньшими словами, -- про параметричне
зображення підмноговиду~$\sigma$, або, ще по-иньшому, про виказ ниті~$\sigma$ з певним відміром змінною~${\xi}$.
Сам підмноговид~$\sigma$ є при цьому {\it слідом} стежки $\tau$. Часом ще говорять про {\it непараметризовані
доріжки,} або ж непараметризовані криві, що є те саме. Якщо стежки, які є розв'язками деякого диференційного
рівняння, ними ж і залишаються при зміні відміру (тобто, при довільному локальному невиродженому перетворенні
незалежної змінної~${\xi}$), то кажемо про відповідне рівняння, що воно {\it нев\'ідмірне} (часом кажуть
„непараметризоване“, або ж „параметрично--байдуже“).

Упохіднення за змінною~${\xi}$ позначаємо крапкою, коваріянтне упохіднення -- штрихом; замість~$\dot x$
пишемо~$u$. Простір змінних $\{{x^n, u^n, \dot u^n\dots\stackrel{\SS{\overbrace{\ldots}^{r-1}}}{u}{}^n}\}$ має
ім'я. Це є простір $r$--швидкостей Ересмана~$T^r\frak M$, який має своє, нутрішнє означення, що не залежить од
впровадженої тим чи иньшим способом системи координат. Функцію Ляґранжа позначаємо літерою~$L$,
$L:\, U\subset{T^r\frak M}\to \Bbb R$. Ту саму функцію, виражену в координатах $\{x^n, u^n, \dot u^n, \ddot
u\dots\}$, позначаємо~$L^u$.
\subsection{В\'ідмірна байдужість у механіці Остроградського.}
На многовиді~$T^r\frak M$ означені так звані „основні ділачі“ $\zeta_{\SS1}\dots\zeta_{r}$, які узагальнюють
поле Ліувіля~\cite{matsyuk:de Leon},~\cite{matsyuk:MKaw}. З їх поміччю можна висловитися, зокрема, щодо умов
безвідмірності ріжних побудов, які творимо на просторі~$T^r\frak M$. До порядку~2 ці ділачі є такими:
\begin{equation}
  \label{matsyuk:FundFields}
\zeta_1=u^n\dfrac{\partial}{\partial u^n}+2\,\dot u^n\dfrac{\partial}{\partial\dot u^n},\qquad
\zeta_2=u^n\dfrac{\partial}{\partial\dot u^n}\,.
\end{equation}

Деяка функція~$f$, задана на (частині) простору~$T^2\frak M$, не є залежною від перетворення похідних $u^n, \dot
u^n$, зумовленого довільним (локальним, вза\-є\-м\-но--однозначним) перетворенням незалежної змінної~$\xi$, в тім,
і тільки, разі, коли
\begin{equation}
  \label{matsyuk:parind}
\zeta_{1}f=0,\qquad\zeta_{2}f=0\,.
\end{equation}

З другого боку, деяка иньша функція~$L$, задана на (частині) простору~$T^2\frak M$, створює параметрично--байдуже
варіяційне завдання з функціоналом $ \int L(x^n,u^n,\dot u^n)\,d\xi $, якщо, і тільки, наступні умови Цермело
задоволені(гляди~\cite{matsyuk:MKaw}, або ще~\cite{matsyuk:Logan}, уклад~8.19):
\begin{equation}\label{matsyuk:Zermelo}
\zeta_1L=L,\qquad\zeta_2L=0\,.
\end{equation}

У прийнятих в механіці Остроградського поняттях кількостей руху
\begin{equation}
  \label{matsyuk:p} p^{(1)}_n=\dfrac{\partial L^u}{\partial\dot u^n},\qquad p_n=\dfrac{\partial L^u}{\partial
u^n}-\frac{dp^{(1)}_n}{d\xi}
\end{equation}
вираз, і відповідне рівняння Ойлера--Пуасона подаються таким утвором:
\begin{equation}\label{matsyuk:E-P}
{\cal E}_n=\dfrac{\partial L^u}{\partial x^n}-\dfrac{d p_n}{d\xi}=0\,,
\end{equation}
а функція Гамільтона таким:
\[
H=p^{(1)}_n\dot u^n+p_nu^n-L^u \,.
\]
Спосіб виказу функції Гамільтона у поняттях ділачів~(\ref{matsyuk:FundFields}) був отриманий в
Україні~\cite{matsyuk:thesis} і, незалежно, в іберійському світі~\cite{maciuk:CRAS}:

\medskip
\begin{PRP}
\[
H=\zeta_1L-\frac{d}{d\xi}\zeta_2L-L\, .
\]
\end{PRP}
Наступне речення стає тепер очевидним:

\medskip
\begin{PRP}\label{matsyuk:Ham}
Якщо деяка функція $L_{{\rm I}}$ творить параметрично--байдуже варіяційне завдання і якщо деяка иньша функція
$L_{\rm II}$ є безвідмірною,  то в такому випадку відповідна функція Гамільтона
$H_{L_{\rm{II}}+L_{\rm{I}}}=-L_{\rm{II}}$, так що функція $L_{\rm{II}}$ є сталою руху для екстремалей варіяційного
завдання з функцією Ляґранжа $L=L_{\mathrm{II}}+L_{\mathrm{I}}$.
\end{PRP}

У двовимірному світі можна добути квадратний корінь з виразу \linebreak$\pr{\pw u{u'}}{\pw u{u'}}$,
який входить до означення кривини ниті:
\begin{equation}
\label{matsyuk:Frenet} k=\dfrac{\nw{\bu}{\bw}}{\N3}=\pm\dg\,\dfrac{\ep k nu^ku'\msp^n}{\N3}\,.
\end{equation}

\medskip
\begin{PRP}\label{PRPlagrin2dimRieman}
Варіяційне завдання з функцією Ляґранжа
\begin{equation}\label{matsyuk:L total in k}
  L^{\cal R}=k-{m}\,\nbu
\end{equation}
творить геодезійні кола у двовимірному (лже)ріманівському многовиді.

\end{PRP}
{\sl Доведення.} Що множина геодезійних кіл вичерпується слідами екстремальних стежок варіяційного
завдання~(\ref{matsyuk:L total in k}), я намагався довести в недавній роботі, яка доступна лиш в
Інтернеті~\cite{maciuk:SIGMA}~$\heartsuit.$

\subsection{Коваріянтність.}\label{subsectionCovariance}
Вираз Ойлера--Пуасона означує добре збудовану нутрішню геометричну сутність співзмінної („коваріянтної“) природи;
кількість же руху~$p_n$ з другого означення в ук\-ла\-ді~(\ref{matsyuk:p}) -- ні. Аби подати вираз Ойлера--Пуасона у
вигляді, який полегшує числення на многовиді, оснащеному засобами співзмінного упохіднення (як-от ріманівською
лучністю), добре було б запровадити співзмінні кількості руху (ще кажуть „імпульси“). Робиться воно ось як.

У просторі~$T^2\frak M$ переводимо заміну координатних функцій:
\[\upsilon\,:\{x^n,u^n,\dot u^n\}\mapsto\{x^n,u^n,u'\msp^n\}
\]
і запроваджуємо позначку: $L^{\upsilon}=L^u\circ\upsilon^{-1}$. Маємо взори для перерахунку похідних:
\begin{equation}\label{from u dot to u'}
    \frac{{\partial
L^u}}{\partial \dot u^n}=\frac{\partial L^{\upsilon}}{\partial  u'\msp^n}\,,\;
    \frac{{\partial
L^u}}{\partial u^n}=\frac{\partial L^{\upsilon}}{\partial u^n}+2\,\frac{\partial L^{\upsilon}}{\partial
u'\msp^q}\ga q m nu^m,\; \frac{{\partial L^u}}{\partial x^n}=\frac{{\partial L^{\upsilon}}}{\partial
x^n}+\frac{\partial L^{\upsilon}}{\partial u'\msp^q}\,\frac{\partial\ga q m l}{\partial x^n}\,u^lu^m.
\end{equation}

Наступним кроком запроваджуємо співзмінні кількості руху
\begin{equation}
 \bq =\dfrac{\partial L^{\upsilon}}{\partial \bw}\,,\qquad
\label{matsyuk:pi1}
 \bp =\dfrac{{\partial L^{\upsilon}}}{\partial \bu}-\bq\bpr \,.
\end{equation}

\medskip
\begin{PRP}
Припустімо, що функція Ляґранжа~$L$ залежить тільки від диференційних інваріянтів $\gamma=\pr uu$, $\beta=\pr
u{u'}$, $\alpha=\pr{u'}{u'}$. У такому випадку вираз Ойлера--Пуасона є:
\begin{equation}
  \label{matsyuk:E II ultimate}
\fbox{$\displaystyle{\cal E}_n=-\pi'{}_n-\pi^{(1)}{}_qR_{nkm}{}^qu^mu^k$}
\end{equation}\end{PRP}

\bigskip\noindent
{\sl Доведення переходить такими кроками:}

\smallskip\noindent
\newcounter{PRPcounter}\newbox\STEP\setbox\STEP\hbox{\sl Крок}
\begin{list}{\sl Крок~\arabic{PRPcounter}.\hfill}{\usecounter{PRPcounter}\leftmargin0pt
\setlength{\itemindent}{\wd\STEP}
}
\item
Співзмінну кількість руху~$\bp$, скориставши з першого співвідношення укладу~(\ref{from u dot to u'}) та взірця,
як підраховувати співзмінну похідну~(\ref{matsyuk:AppCov}), подаємо так:
\begin{equation}
{\pi_n} =\dfrac{\partial L^u}{\partial u^n}-2\,\ga q m n u^m \pi^{(1)}{}_q
 -{\pi^{(1)}}\msp'{}_n\,.
 \label{matsyuk:pi}
\end{equation}
Співзмінну похідну від кількості руху~$\bq$, знову ж, користаючи із взірця~(\ref{matsyuk:AppCov}), записуємо ось
як:
\begin{equation} \label{matsyuk:pi1'}
  {\pi^{(1)}}\msp'{}_n=\frac d{d\xi}\pi^{(1)}{}_n-\ga m l n \pi^{(1)}{}_m u^l \,.
\end{equation}
\item
Посередництвом повищих співзмінних величин, змінна~$p_n$ із укладу~(\ref{matsyuk:p}) виражається так:
\begin{eqnarray}
  p_n  &=  & \pi_n+2\,\ga q m n u^m\pi^{(1)}{}_q+\pi^{(1)}\msp'{}_n-\phantom{\frac d{d\xi}\pi^{(1)}{}_n}\quad\mbox{\rm (на основі (\ref{matsyuk:pi}))} \nonumber \\
   &  & \phantom{\pi_n+2\,\ga q m n u^m\pi^{(1)}{}_q+\pi^{(1)}\msp'{}_n}-\frac d{d\xi}\pi^{(1)}{}_n \quad\mbox{\rm (на основі (\ref{matsyuk:pi1}))}
   \nonumber\\[2\jot]
   & = &\pi_n+\ga q m n u^m\pi^{(1)}_q \qquad\qquad\mbox{\rm (на основі (\ref{matsyuk:pi1'})).}\label{matsyuk:p_n}
\end{eqnarray}
\item
Упохіднюючи~(\ref{matsyuk:p_n}) і застосовуючи~(\ref{matsyuk:AppCov}) для того, щоб виразити звичайну похідну від
змінних $\pi$ та $u$ в термінах, відповідно, співзмінних похідних $\bp\bpr$ та $\bu\bpr$, разом
з~(\ref{matsyuk:pi1'}), отримуємо:
\begin{eqnarray*}
 \frac d{d\xi} p_n&=&\pi'{}_n+\ga l m n \pi_l u^m+\dfrac{\partial \ga l
m n}{\partial x^k}
 u^ku^m\pi^{(1)}{}_l \\
&& {}+\big(\ga l m n {u'}\msp^m-\ga l m n
 \ga m q k u^qu^k\big)\pi^{(1)}{}_l
 +\ga l m n u^m\big({\pi^{(1)}}\msp'{}_l
 +\ga q k l \pi^{(1)}{}_q u^k\big)  \\[2\jot]
&=&\pi'{}_n+\big(\pi^{(1)}\msp'{}_l+\pi_l\big)\ga l m n u^m+\pi^{(1)}{}_l\ga l m n u'\msp^m \\[2\jot]
&&{}+\pi^{(1)}{}_q u^m u^k\left(\ga l m n \ga q l k+\dfrac{\partial \ga q m n}{\partial x^k}
        -\ga q l n \ga l m k\right).
\end{eqnarray*}
\item
Вираз Ойлера--Пуасона~(\ref{matsyuk:E-P}) тепер прибирає вид:
\[
{\cal E}_n=\dfrac{\partial L^{\upsilon}}{\partial x^n}
        -\big(\pi^{(1)}\msp'{}_l+\pi_l\big)\,\ga l m n u^m
        -\pi^{(1)}{}_l\ga l m n u'\msp^m
        -\pi'{}_n-\pi^{(1)}{}_l u^mu^kR_{nkm}{}^l \,.
\]
Покажемо, що перші чотири доданки в цьому виразі творять зеро,~-- в умовах Речі, яку доводимо. Задля складання
виразу
\begin{equation}\label{maciuk:*}
\frac{\partial L^{\upsilon}}{\partial x^n}=\frac{\partial L^{\upsilon}}{\partial\gamma}\frac{\partial
\gamma}{\partial x^n}+\frac{\partial L^{\upsilon}}{\partial\beta}\frac{\partial\beta}{\partial x^n}+\frac{\partial
L^{\upsilon}}{\partial\alpha}\frac{\partial\alpha}{\partial x^n}\,,
\end{equation}
підрахуємо, послужившись взором~(\ref{matsyuk:AppDxG-1}):\[\frac{\partial \gamma}{\partial x^n}=2\ga
lmnu^mu_l\,,\quad\frac{\partial \beta}{\partial x^n}=\ga lmnu^mu'{}_l+\ga lmnu'\msp^mu_l\,,\quad\frac{\partial
\alpha}{\partial x^n}=2\ga lmnu'\msp^mu'{}_l\,.\] З другого боку, виходячи з означення~(\ref{matsyuk:pi1}), маємо:
\begin{equation}\label{maciuk:**}\pi^{(1)}{}_n=\frac{\partial L^{\upsilon}}{\partial \beta}\,u_n+2\frac{\partial L^{\upsilon}}{\partial \alpha}\,u'{}_n\,,\quad
\pi^{(1)}\msp'{}_n+\pi_n=2\,\frac{\partial L^{\upsilon}}{\partial \gamma}\,u_n+\frac{\partial L^{\upsilon}}{\partial
\beta}\,u'{}_n\,.
\end{equation}
Віднявши вирази~(\ref{maciuk:**}), з відповідними множниками $\ga l m nu'\msp^m$ та $\ga l m nu^m$, од виразу~(\ref{maciuk:*}), одержуємо зеро~$\heartsuit.$
\end{list}
\subsection{Рівняння Ойлера--Пуасона для означеної кривини.}
Візьмімо за функцію Ляґранжа, про яку йдеться в Речі~\ref{matsyuk:Ham}, так звану {\it означену} кривину
\begin{equation}
  \label{matsyuk:L II}
L_{\rm II}=\tilde k\stackrel{\mbox{\tiny def}}=\dg\,\frac{\ep m nu^mu'\msp^n}{\N3}\, .
\end{equation}
Для такої функції Ляґранжа співзмінні імпульси суть:
\begin{equation}
  \label{matsyuk:Pi1}
\pi^{(1)}{}_n=\dg\,\frac{\ep k nu^k}{\N3}\,, \qquad
\pi_n=\dg\,\frac{\ep k nu'\msp^k}{\N3}\,,
\end{equation}
Докладніше про техніку здобуття виразів~(\ref{matsyuk:Pi1}) можна довідатися з
інтернет-письма~\cite{maciuk:SIGMA}.

Підрахуймо похідну~$\pi'$ від кількості руху~$\pi$ в укладі~(\ref{matsyuk:E II ultimate}). При цьому важливо
пам'ятати, що співзмінну похідну~$\pi'{}_n$ потрібно обчислювати, виходячи з означення~(\ref{matsyuk:AppCov}),
застосованого до другого виразу в укладі~(\ref{matsyuk:Pi1}), і, так само, похідну~$\frac {du'\msp^k}{d\xi}$
потрібно обчислювати, виходячи з того ж правила~(\ref{matsyuk:AppCov}):
\[
\frac{du'\msp^k}{d\xi}=u''\msp^k-\ga k l mu'\msp^mu^l.
\]
Користаючи з укладу (\ref{matsyuk:AppDxg}), врахувавши~(\ref{matsyuk:AppModu'}) разом зі взором
спрощення~(\ref{matsyuk:g-simpl}), отримаємо
\[
\pi'_n=-3\,\dfrac{\dg(\pr u{u'})}{\nbu^5}\,\ep k nu'\msp^k
        +\dfrac{\dg}{\nbu^3}\,\ep k nu''\msp^k.
\]

\medskip
\begin{PRP}
У двовимірному світі рівнянням Ойлера--Пуасона для функції Ляґранжа
\begin{equation}\label{matsyuk:L total}
\tilde L^{\cal R}=\tilde k-{m}\,\nbu
\end{equation}
є:
\begin{eqnarray}\label{equationinRieman}
\fbox{$\displaystyle  {m}\,\dfrac{(\pr u u)\,u'{}_n-(\pr{u'}u)\,u_n}{\nbu^3} = \dfrac{\om k nu''\msp^k}{\N3}
       -3\,\dfrac{(\pr u{u'})}{\N5}\,\om k nu'\msp^k
+
   \dfrac{K}{\nbu}\,\om n mu^m \label{matsyuk:E alt-1}$}
\end{eqnarray}
де~$K=R_{12}{}^{12}$ є кривина многовиду, $\omega_{nm}=\dg\ep n m$ є елемент об'єму.
\end{PRP}
{\sl Доведення.} Все, що осталося зробити~-- безпосереднім обчисленням отримати наступне спрощення у правій часті
взору~(\ref{matsyuk:E II ultimate}):
\begin{equation}\label{cal R}
  {\cal R}_n\stackrel{\mbox{\tiny def}}=\dfrac{\dg}{\N3}\,\ep m qR_{nkl}{}^qu^mu^lu^k=\dfrac{\dg K}{\nbu}\,\ep n
  mu^m\qquad\heartsuit.
\end{equation}

\medskip
\begin{REM}\label{maciuk:k'=0}
Рівняння~(\ref{matsyuk:E alt-1}), стяте зі швидкістю~$u^n$, має вигляд \[0=-\frac {d\tilde k}{d\xi}\,,\] що
безпосередньо ще раз підтверджує сталість кривини екстремальної ниті.
\end{REM}
\paragraph{\fontseries{bx}\fontshape{it}\selectfont Діярій.} Річ~\ref{PRPlagrin2dimRieman} давно була мені
відома~\cite{matsyuk:1981}, як вислід заплутаних по-ко\-ор\-ди\-нат\-них підрахунків, виснажливих своєю
безнадійно-одноманітною невпорядкованістю. Властивість кожної з кривин Френе бути сталою на естремальних нитях
такого варіяційного завдання, де сама вона виступає функцією Ляґранжа, є одним з наслідків Речі~\ref{matsyuk:Ham},
але я цю властивість усвідомлював тільки в рамцях пласк\'ого простору (гляди~~\cite{matsyuk:thesis}). Причина моєї
короткозорості полягала в тім, що сама Річ~\ref{matsyuk:Ham} до останнього часу успішно від мене ховалася. В
письмі~\cite{matsyuk:1982} згадана властивість щодо першої кривини Френе зостала голослівно поширена і на
викривлений простір наслідком простої описки на сторінці~87, не зауваженої при вичитуванні. Сам\'е варіяційне
рівняння~(\ref{equationinRieman}) було відомим тільки в пласк\'ому просторі, як єдиний розв'язок певного
симетрійного оберненого варіяційного завдання~\cite{matsyuk:1982}. Вигляд рівняння~(\ref{equationinRieman}) в
загальному ріманівському оточенні остававсь таємницею з тієї причини, що „коваріянтні“ міркування
Відділу~\ref{subsectionCovariance}, зокрема, поняття співзмінних кількостей руху та їх виразів~(\ref{matsyuk:Pi1})
ніяк не вирізнялися у мряці сумно згаданих по-координатних обчислень.

\section{РІВНЯННЯ ДІКСОНА І ВАРІЯЦІЙНЕ УЗАГАЛЬНЕННЯ РІВНЯНЬ МАТІСОНА}\label{Dixon}
Рівняння Матісона~(\ref{matsyuk:5}) можна отримати з більш загальної системи {\it рівнянь
Діксона}~\cite{matsyuk:Dixon}
\begin{eqnarray}\label{maciuk:1}
  P' & = & F\,, \\
  S' & = & P\wedge u\,,    \label{maciuk:2}
\end{eqnarray}
де величина $P$ є деяким вектором, позначка $P\wedge u$ означає скісний тензор, збудований за правилом
$(P\wedge u)_{nm}=P_{n}u_{m}-P_{m}u_{n}$, а сила взаємодії дзиґи~$S$ із ріманівською кривиною~$R_{nm}{}^{kl}$ є
такою{\renewcommand{\thefootnote}{\fnsymbol{footnote}}\footnote[1]{Наше означення тензора
кривини~(\ref{matsyuk:AppR}) ріжниться знаком від означення, прийнятого Діксоном.}}:
\begin{equation}\label{maciuk:F}
  F_n=-{1\over2}R_{nm}{}^{kl}u^mS_{kl}\,.
\end{equation}
Виведення рівняння~(\ref{matsyuk:5}) із системи рівнянь~(\ref{maciuk:1}, \ref{maciuk:2}) в чотиривимірному світі
супроводжується накладанням додаткової умови Матісона--Пірані~(\ref{matsyuk:MP}) разом з її диференційним
наслідком -- {\it упохідненою умовою Матісона--Пірані\/} (гляди \cite{matsyuk:4}, уклад~(3.21))
\begin{equation}\label{maciuk:3}
  S'u+Su'=0\,.
\end{equation}
У двовимірному світі будь-яка додаткова умова є непотрібною, оскільки матриця скісного тензора~$S$ невироджена, і,
відповідно, рівнянь (\ref{maciuk:1},~\ref{maciuk:2}) достатньо для визначення світової ниті. Ці рівняння є
невідмірними: з них можна довідатися тільки про непараметризовані доріжки. Часом, для більшої визначеності та
полегшення рахунків, до таких невідмірних рівнянь додають іще одне, наприклад, рівняння~(\ref{u2=1}), яке усталює
відмір. Тоді доповнена система рівнянь робиться геть невиродженою. Як зазначалося у Вступі, рівняння
Матісона~(\ref{matsyuk:5}) вже є відміряне власним часом~(\ref{u2=1}). Такий вибір відміру нас не влаштовує. Бо у
нашім випадку він несумісний з упохідненою умовою Матісона--Пірані. Покажемо, чому.

У двовимірному світі скісний тензор~$S_{nm}$ може вважатися співвідносним з елементом об'єму:
\begin{equation}\label{maciuk:f}
S=-f\omega.
\end{equation}
Відповідно, $S'=-f'\omega$. Упохіднена умова Матісона--Пірані має наслідком рівняння колінеарности
\begin{equation}\label{u^u'}
   u'\wedge u=0\,,
\end{equation}
і, властиво, зводиться до вибору множника пропорційності $-f'/f$ в ньому. В'язь же~(\ref{u2=1}), як і взагалі,
вимога $\pr u u=\rm const$ тепер узгодиться з упохідненою умовою Матісона--Пірані хіба що тільки у благ\'ому
випадку~$f'=0$. Упохіднена умова Матісона--Пірані є більш ліберальною: рівняння колінеарности, забороняючи,
правда, прискорення~$k$, не вимагає при цьому постійности {\it крутня} (просторово--од\-но\-ви\-мір\-но\-го
„спіну“). Таким чином, можемо замінити спосіб відміру, прийнятий у рівняннях Матісона, на иньший спосіб відміру
світової ниті,-- приєднавши до рівнянь Діксона~(\ref{maciuk:1},~\ref{maciuk:2}) упохіднену умову
Ма\-ті\-со\-на--Пі\-ра\-ні~(\ref{maciuk:3}). Звісно, про вигляд рівнянь Матісона~(\ref{matsyuk:5}) прийдеться
забути, подібно до того, як ми розпрощалися з „інтеґральною“ умовою Матісона--Пірані~(\ref{matsyuk:MP}).

В попередньому дописі (уклад~16 праці~\cite{NTSH2006}) пропонувалася така видозміна рівняння Матісона, яка вільна
од умови $\pr u u=1$ і виведена в ширших рамцях тільки лиш упохідненої умови Матісона--Пірані:
\begin{equation}\label{maciuk:16}
\fbox{$\displaystyle  M\,\frac{\uu u-\N2u'}{\N3}=3\frac{\uu}{\N4}Su'-\frac{Su''}{\N2}-F$}
\end{equation}
Лівий бік цього рівняння є виразом Ойлера--Пуасона для функції Ляґранжа~$M\nbu$, якщо тільки вважати величину~$M$
сталим числом.

Можна було б і не долучати жодних додаткових умов відміру до рівнянь Діксона -- ці рівняння й так уже самі собою
визначать світову нить. Та нам цікаво, чи  можливе варіяційне вираження системи Діксона рівноцінним рівнянням без
змінної $S$, зате вищого (третього) порядку. Тут важливим є факт, що у двовимірі варіяційне рівняння третього
порядку ніяк не може бути параметрично-байдужим, оскільки у варіяційному рівнянні третього порядку біля найстаршої
похідної множником завжди мусить стояти скісна матриця, яка у двовимірі є невиродженою. Иньшими словами, шукаємо
варіяційне рівняння третього порядку, „гамільтонізація“ якого виллється у систему рівнянь Діксона.

\subsection{Рівняння Діксона і рівняння Матісона у двовимірі.}
Для узасаднення рівняння~(\ref{maciuk:16}) придивімось ближче до системи рівнянь
(\ref{maciuk:1},~\ref{maciuk:2},~\ref{maciuk:3}).

\medskip
\begin{PRP}\label{maciuk:PRP1}У двовимірі рівняння~(\ref{maciuk:2}) алгебрично рівносильне з таким:
\begin{equation}\label{maciuk:2P}
  P=\frac{Pu}{\pr uu}u+\frac{S'u}{\pr uu}.
\end{equation}
\end{PRP}

\medskip
Для доведення складімо скісний добуток виразу~(\ref{maciuk:2P}) з вектором швидкості~$u$, залучивши виказ змінної
„спіну“~(\ref{maciuk:f}) та скориставшись взором спрощення~(\ref{maciuk:Gajda 5_A.3}):
\begin{eqnarray*}
  (\pr uu)\pw Pu_{nm} & = & \SP nku^ku_m-\SP mku^ku_n=\dg f'\left(\ep nku^ku_m-\ep mku^ku_n\right) \\
  & = & (\pr uu)\dg f'\ep nm=(\pr uu)\SP nm \qquad\heartsuit.
\end{eqnarray*}

\medskip\noindent
Долучімо до системи рівнянь (\ref{maciuk:1},~\ref{maciuk:2},~\ref{maciuk:3}) іще дві величини, $M$ та $\overline
M$, разом з двома рівняннями, що їх означують:

\medskip
\begin{eqnarray}\label{maciuk:M}
  M & = & \frac{Pu}{\nbu} \,,\\
  \overline M & = &  \frac {P'u}{\nbu}+\frac {Pu'}{\nbu}-\frac {\uu}{\N3}Pu\,.     \label{maciuk:12'}
\end{eqnarray}

\medskip
Рівняння
\begin{equation}\label{maciuk:12}
  \overline M=0
\end{equation}
-- це тотожна в'язь, яка є алгебричним наслідком системи (\ref{maciuk:1},~\ref{maciuk:3},~\ref{maciuk:2P}). Щоб у
цьому переконатися, досить стяти рівняння~(\ref{maciuk:1}) зі швидкістю~$u$, а рівняння~(\ref{maciuk:2P}) -- з
похідною від швидкості, $u'$, пам'ятаючи про скісну симетрію тензорів кривини та „спіну“ та скориставши з
переметної властивості~(\ref{maciuk:3})~{$\heartsuit.$}\endgraf

\medskip\noindent
Величина $\overline M$ сприймається як похідна від величини~$M$.

\medskip
Тепер на многовиді, заданому рівняннями
(\ref{maciuk:3}) та (\ref{maciuk:M}), виказ кількості руху~(\ref{maciuk:2P}) можна записати в друг\'ому
вигляді:

\medskip
\begin{equation}\label{maciuk:4}
  P=\frac M{\nbu}u-\frac{Su'}{\N2}\,.
\end{equation}

\newpage
\bigskip
Рівняння~(\ref{maciuk:2}), стяте з похідною~$u'$, з використанням рівняння~(\ref{maciuk:4}), творить ще один
наслідок:
\begin{equation}\label{maciuk:160}
  S'u'+\frac{\uu}{\pr uu}Su'=0.
\end{equation}
Все ще перебуваючи на многовиді, заданому рівняннями (\ref{maciuk:3}) та (\ref{maciuk:M}), ще раз долучімо до
системи рівнянь (\ref{maciuk:1},~\ref{maciuk:2}) деяку нову величину~$\overline P$ разом з рівністю, що її
означує:
\begin{equation}\label{maciuk:15}
  \overline P=M\biggl(\frac u{\|u\|}\biggr)'+3\frac{\uu}{(\pr uu){\SS2}}Su'-\frac{Su''}{\pr uu}\,.
\end{equation}
З огляду на властивість~(\ref{maciuk:160}), разом з властивістю~(\ref{maciuk:12}), всюди на
многовиді~(\ref{maciuk:3}) нову величину~$\overline P$ можна сприймати, в рамцях системи
(\ref{maciuk:1},~\ref{maciuk:2}), як похідну від кількості руху~(\ref{maciuk:4}):
\begin{equation}\label{P'-Q}
  P'-\overline P=0\,.
\end{equation}
Безпосереднім алгебричним наслідком рівняння~(\ref{maciuk:1}), враховуючи (\ref{P'-Q}) і ~(\ref{maciuk:15}), є
рівняння~(\ref{maciuk:16}).

\medskip
\begin{PRP}\label{maciuk:PRP2}
На многовиді, заданому рівняннями ~(\ref{maciuk:3}), ~(\ref{maciuk:M}), ~(\ref{maciuk:12}), ~(\ref{maciuk:4}),
~(\ref{maciuk:15}), ~(\ref{P'-Q}), рівняння~(\ref{maciuk:16}) є алгебрично рівноцінним із системою рівнянь
(\ref{maciuk:1},~\ref{maciuk:2}).
\end{PRP}
{\sl Доведення.} Щойно було вказано, як рівняння~(\ref{maciuk:16}) випливає із системи Діксона. Навпаки, це ж саме
рівняння~(\ref{maciuk:16}), в термінах величини~$\overline P$ з означення~(\ref{maciuk:15}), записується як
\begin{equation} \nonumber
  \overline P=F\,,
\end{equation}
що чинить утвір~(\ref{maciuk:1}) в ототожненні~(\ref{P'-Q})~{$\heartsuit.$}\endgraf

\medskip
\begin{CMM}
\rm Вираз~(\ref{maciuk:4}) запроваджує змінну кількості руху~$P$ замість похідної од швидкості,~$u'$, яка присутня
у рівнянні~(\ref{maciuk:16}). Виразом~(\ref{maciuk:15}) запроваджується ще одна змінна,~$\overline P$, замість
другої похідної від швидкості,~$u''$, що теж присутня в рівнянні~(\ref{maciuk:16}). Ця процедура лежить в рамцях
ідеології переписування рівнянь з вищими похідними в термінах рівнянь першого порядку, до якої, зокрема,
відноситься й спосіб гамільтонізації в механіці Остроградського.
\end{CMM}

\medskip
\begin{CMM}
\rm Уклад~(\ref{P'-Q}) надає цікаву інформацію: він вказує на те, що для встановлення рівноцінності
рівняння~(\ref{maciuk:16}) зі системою Діксона (\ref{maciuk:1},~\ref{maciuk:2}), яка є першого порядку, необхідно
накинути на один порядок вищу в'язь, яка, виходячи з Речі~\ref{maciuk:PRP1} та виразу кількості
руху~(\ref{maciuk:4}), є рівнозначною з упохідненням рівняння~(\ref{maciuk:2}).
\end{CMM}

\medskip
\begin{CMM}
\rm Укладом~(\ref{maciuk:12}) вимагається збереження деякої комбінації, яка впроваджується
виразом~(\ref{maciuk:M}). Ця вимога є додатковою до рівняння~(\ref{maciuk:16}).
\end{CMM}

\medskip
\begin{CMM}
\rm Ми ніде не використовували „інтеґральну“ умову Матісона--Пірані, а тільки упохіднену від неї
умову~(\ref{maciuk:3}).
\end{CMM}

\medskip
\begin{CMM}
\rm Міркування, які завели до Речі~\ref{maciuk:PRP2}, дослівно переносяться в чотиривимірний світ. Єдина ріжниця
полягає у виказі Речі~\ref{maciuk:PRP1}: там, поруч з рівнянням~(\ref{maciuk:2P}), з'являється додаткове рівняння
$\epsilon_{nmkl}u^mS'\msp^{kl}=0$, природа якого пов'язана зі збільшенням свободи в
системі~(\ref{maciuk:1},~\ref{maciuk:2}), коли її розглядати в чотиривимірі (гляди~\cite{NTSH2006}, уклад~9).
\end{CMM}

\medskip
\begin{PRP}
Нехай в „рівнянні Матісона“~(\ref{maciuk:16})
\begin{equation}\label{maciuk:S}
S=-\frac{\omega}{\nbu}\,,
\end{equation}
де $\omega$ є елементом об'єму. В цьому разі рівняння~(\ref{maciuk:16}) перетворюється в рівняння Ойлера--Пуасона
для функції Ляґранжа~(\ref{matsyuk:L total}).
\end{PRP}
{\sl Доведення.} Порівнюючи рівняння~(\ref{equationinRieman}), вкупі зі співвідношенням~(\ref{cal R}), із
рівнянням~(\ref{maciuk:16}), вкупі з означенням~(\ref{maciuk:F}), досить показати, що
\begin{equation}\label{F=R}
  F_n=\frac{\dg}{\N3}\ep m qR_{nkl}{}^{q}u^mu^lu^k\,,
\end{equation}
якщо у виразі~(\ref{maciuk:F}) замінити тензор крутня виразом~(\ref{maciuk:S}). Иньшими словами, слід переконатися
у справедливості співвідношення
\begin{equation}\nonumber
  ({\pr uu})R_{nq}{}^{kl}u^q\ep kl=2\ep mqR_{nkl}{}^qu^mu^lu^k\,.
\end{equation}
Виходячи зі взору спрощення~(\ref{maciuk:Gajda 5_A.3}), лівий бік повищої рівності, завдячуючи скісній симетрії
тензора кривини, перетворимо так:
\begin{eqnarray*}
  ({\pr uu})\,R_{nq}{}^{kl}u^q\ep kl & = & R_{nq}{}^{kl}u^qu_k\ep mlu^m - R_{nq}{}^{kl}u^qu_l\ep mku^m \\
   & = & R_{nq}{}^{kl}u^qu_k\ep mlu^m + R_{nq}{}^{lk}u^qu_l\ep mku^m \\
   & = & 2\ep mlR_{nqk}{}^{l}u^mu^qu^k\qquad\heartsuit.
\end{eqnarray*}
\medskip
\subsection{Імпульсний виказ.}
Систему рівнянь Діксона (\ref{maciuk:1},~\ref{maciuk:2}) можна отримати безпосередньо із виразу
Ойлера--Пуасона~(\ref{matsyuk:E II ultimate}), скориставшись, як означеннями, виразами для кількості руху~$\bp$ та
вищого імпульсу~$\bq$ з укладу~(\ref{matsyuk:Pi1}). Дійсно, із співвідношень (\ref{maciuk:F}) та (\ref{F=R}), при
запровадженні тензора крутня взором~(\ref{maciuk:S}), одразу бачимо, що рівняння Ойлера--Пуасона~(\ref{matsyuk:E
II ultimate}) набуває вигляду~(\ref{maciuk:1}), якщо покласти
\begin{equation}
  P=-\bp\,.
\end{equation}

Що ж до підрахунку виразу $P\wedge u$, скористаємося наслідком взору спрощення~(\ref{p.G}), \[\uu\, \ep nm -
u_n\ep kmu'^k + u_m\ep knu'^k =0\,,\] в такий спосіб:
\begin{equation}\nonumber
  \left(u\wedge \bp\right)_{nm}=\frac{\dg}{\N3}\left(\ep kmu'\msp^ku_n-\ep
  knu'\msp^ku_m\right)=\frac{\dg}{\N3}\uu\,
  \ep nm=\SP nm\,,
\end{equation}
що творить рівняння~(\ref{maciuk:2}).
\begin{REM}
Тензор „спіну“~(\ref{maciuk:S}) можна виразити через поняття вищого імпульсу~$\bq$:
\begin{equation}\nonumber
  S=\bq\wedge u\,.
\end{equation}
\end{REM}
Це випливає безпосередньо з означень (\ref{maciuk:S}), (\ref{matsyuk:Pi1}), та взору спрощення~(\ref{maciuk:Gajda
5_A.3})~{$\heartsuit$.}\endgraf
\medskip
\begin{REM}\label{Pu=-piu}
Величина $Pu=-\bp u$ є нічим иньшим, як означеною кривиною~$\tilde k$ із утвору~(\ref{matsyuk:L II}), яка, у свою
чергу, є сталою руху для рівняння~(\ref{matsyuk:E II ultimate}), якщо тільки кількості руху задати відповідними
виразами з укладу~(\ref{matsyuk:Pi1}).
\end{REM}
Це видко із другого виразу в укладі~(\ref{matsyuk:Pi1})~{$\heartsuit$.}\endgraf
\medskip
Погляньмо тепер, як веде себе величина~$M=\frac{Pu}{\nbu}$ з укладу~(\ref{maciuk:M}) на розв'язках
рівняння~(\ref{equationinRieman}), яке, у свою чергу, можна розглядати, як самостійно, так і разом з упохідненою
умовою Матісона--Пірані~(\ref{maciuk:3}). Віднаходимо таке:
\begin{REM}
\item\begin{enumerate}
\item \label{maciuk:item1}На в'язі, яка задана упохідненою умовою Матісона--Пірані~(\ref{maciuk:3}), величина~(\ref{maciuk:M}) є інтеґралом
рівняння~(\ref{equationinRieman}).
\item \label{maciuk:item2}Величина~(\ref{maciuk:M}) є сталою на розв'язках рівняння~(\ref{equationinRieman}) тільки
якщо~$k=0$, або якщо
\begin{equation}\label{maciuk:conditionK}
  \mbox{sgn}(g)\,k^{\SS2}-\sqrt{2}\,\frac{m\tilde k}{\|S\|}-K=0
\end{equation}
(де величина~$\|S\|=\sqrt{|S_{nm}S^{nm}|}$ теж є постійною).
\end{enumerate}
\end{REM}
{\sl Доведення.} Упохіднивши~(\ref{maciuk:M}), з огляду на Зауваження \ref{Pu=-piu} та \ref{maciuk:k'=0}, одержимо
\[M'=-\frac{\uu}{\N3}\tilde k\,,\] так що пункт~\ref{maciuk:item1} вже випливає з рівняння в'язі~(\ref{u^u'}), коли
пригадати собі означення кривини~(\ref{matsyuk:Frenet}).

Аби довести пункт~\ref{maciuk:item2} в часті, коли~$k\neq0$, але й надалі~$M'=0$, спочатку розв'яжімо
рівняння~(\ref{equationinRieman}) щодо найстаршої похідної:
\begin{eqnarray*}
        \dg\left(\dfrac{u''\msp^k}{\N3}
        -3\,\dfrac{(\pr u{u'})\,u'\msp^k}{\N5}\right)
        =m\,\dfrac{(\pr u u)\,e^{nk}u'{}_n-(\pr{u'}u)\,e^{nk}u_n}{\N3}
        -e^{nk}{\cal R}_{n}\,,
\end{eqnarray*}
де $e^{mn}=\det[g_{pq}]\,g^{mk}g^{nl}\ep k l$ означає протизмінний скісний символ Леві--Чивіти, а сила~$\cal R$
означена виразом~(\ref{cal R}). Після стинання зі швидкістю~$u$, беручи під увагу переметне правило~$\pr
{u'}{u'}=-\pr u{u''}$, яке є упохідненою умовою~$\pr u{u'^{\mathstrut}}=0$, отримуємо
\begin{equation}\label{maciuk:u.E alt}
  \dg\,(\pr{u'}{u'})+m\,(\pr uu)\,e^{nk}u'{}_nu_k-\dg(\pr uu)^2K=0\,.
\end{equation}
 Згідно з виразом для означеної кривини стежки~(\ref{matsyuk:L II}), числимо: \[\tilde
k^{2}=\mbox{sgn}(g)\frac{(\pr uu)(\pr{u'}{u'})-(\pr u{u'})^{2}}{\N6}=\mbox{sgn}(g)\frac{(\pr
uu)(\pr{u'}{u'})}{\N6}\,,\] так що остаточно уклад~(\ref{maciuk:u.E alt}) прибирає такий вид:
\[k^{2}-m\tilde k\,\nbu-\mbox{sgn}(g)K=0\,.\] Тепер умова~(\ref{maciuk:conditionK}) випливає зі способу
впровадження змінної крутня~$S_{nm}$ укладом~(\ref{maciuk:S})~$\heartsuit.$
\section{{\rm\it ДОДАТОК:} КОРИСНІ ВЗОРИ.}
\paragraph{У двовимірі справедливими є наступні взори спрощення:}
\begin{eqnarray}
    &\label{p.G}  g_{mn}\ep lk -  g_{lm}\ep nk + g_{km}\ep nl =0\,,&
\\
    \label{matsyuk:e-simpl}
       & g_{(mn)}\ep l k -g_{\,l\,(m}\ep {n)\,} k+g_{\,k\,(m}\ep {n)\,}l =0 \,,&
\\
    \label{maciuk:Gajda 5_A.3}
 & (\pr uu)\,\ep nm-u_n\ep kmu^k+u_m\ep knu^k=0\,,&
\\
     \label{matsyuk:g-simpl}
       & \ep m n \ga l l k - \ep l n \ga l m k + \ep l m \ga l n k=0 \,.&
\end{eqnarray}
Взори (\ref{p.G}) і (\ref{matsyuk:e-simpl}) рівноцінні: антисиметризація взору~(\ref{matsyuk:e-simpl}) вздовж
значків $n,l,k$ творить взір~(\ref{p.G}) з множником~$1/3$.

\paragraph{У (лже)ріманівському просторі є відомими такі уклади:}
\begin{eqnarray}\label{matsyuk:AppCov}
        &\displaystyle a'\,^n=\frac{da^n}{d\xi}+\ga n l m a^m u^l , \qquad a'{}_n=\frac {da_n}{d\xi}-\ga m l n a_m u^l \,,&\\
\label{matsyuk:AppDxG-1}
        &\displaystyle \dfrac{\partial g_{mn}}{\partial x^k}=g_{ml}\ga l k n +g_{nl}\ga l k m \,,&\\
\label{matsyuk:AppDxg}
        &\displaystyle \dfrac{\partial }{\partial x^n}\dg=\dg\ga l l n,\qquad{\rm де}\quad g=\det[g_{nm}]\,,&\\
\label{matsyuk:AppModu'}
        &\displaystyle \left(\dfrac{1}{\strut\N3}\right)'=-3\,\dfrac{\bu\bcdot\bw}{\N5} \,,&\\
\label{matsyuk:AppR}
        &\displaystyle R_{kmn}{}^l=\dfrac{\partial \ga l k n}{\partial x^m}
        -\dfrac{\partial \ga l m n}{\partial x^k}+\ga l m q \ga q k n
        -\ga l k q \ga q m n \,.&
\end{eqnarray}
}

\medskip
Праця виконувалася за підтримки ґранту GA\v CR 201/09/0981 Чеської наукової фундації.
{\newcommand{\ms}{\kern.2em}

}



\bigskip

\begin{center}
{\bf VARIATIONALITY WITH SECOND DERIVATIVES, RELATIVISTIC UNIFORM
 ACCELERATION, AND THE ``SPIN''--CURVATURE INTERACTION IN TWO--DIMENSIONAL SPACE-TIME}

\bigskip

{\it Roman~MATSYUK}

\medskip
Institute for Applied Problems in Mechanics and Mathematics\\3$^{\mbox b}$~Naukova~St., L\kern-.25em'viv, Ukraine

\end{center}

\medskip

A variational formulation for the geodesic circles in two--dimensional Riemannian manifold is discovered. Some
relations with the uniform relativistic acceleration and the one--dimensional ``spin''--curvature interaction is
investigated.


\begin{thebibliography}{99}
\bibitem{HILL1945}
    {\it Hill\ms E.\ms L.,} Phys. Rev. 1947. {\bf 72}. N\ms2. 143--149.
\bibitem{matsyuk:Yano131}
    {\it Yano\ms K.,} Proc. Imp. Acad. Jap. 1940. {\bf 16}. 195--200.
\bibitem{matsyuk:4}
    {\it Mathisson\ms M.,} Acta Phys. Polon. 1937. {\bf 6}. Fasc.\ms3. 163--200.
\bibitem{maciuk:Ragusa}
    {\it Ragusa\ms S., Bailyn\ms M.,} Gen. Relat. Gravitation. 1995. {\bf 27.} No.\ms 2. 163--169.
\bibitem{maciuk:Plyatsko}
    {\it Plyatsko\ms R., Stefanyshyn\ms O.,} Acta. Phys. Pol.~B. 2008. {\it 39.} Fasc.\ms1. 23--24.
\bibitem{maciuk:Natario}{\it Nat\'ario\ms J.,} Comm. Math. Phys. 2008. {\bf 281.} 387--400.
\bibitem{matsyuk:de Leon}{\it de L\'eon\ms M., Rodrigues\ms P.\ms R.,} Generalized
        classical mechanics and field theory.~--  Amsterdam, Elsevier, 1985, xvi,~290\ms p.
\bibitem{matsyuk:MKaw}
        {\it Kawaguchi\ms M.,} RAAG Memoirs. 1968. {\bf 4}. 578--592 (Misc. {\bf 6}. 86--100).

\bibitem{matsyuk:Logan}{\it Logan\ms J.\ms D.,} Invariant variational principles.~-- New York, Academic Press, 1977, xvi,~172\ms p.
\bibitem{matsyuk:thesis}
    {\it Мацюк\ms Р.,} Пуанкаре-инвариантные уравнения движения в лагранжевой механике с высшими производными. дис.\ldots к-та
    физ.­мат. наук. Львов,~1984.~140\ms с.
\bibitem{maciuk:CRAS}
    {\it de Le\'on\ms M, Rodrigues\ms P.\ms R.,} C.\ms R.\ms Acad. Sc. Paris. S\'er.\ms II. 1985. {\bf 301}. n\kern-.25em$^{\scriptscriptstyle\rm
    o}$\ms7. 455--458.
\bibitem{maciuk:SIGMA}
    {\it Matsyuk\ms R.\ms Ya.,} SIGMA. 2008. {\bf 4}. 016. DOI: 10.3842/SIGMA.2008.016. http://www.emis.de/journals/SIGMA/
\bibitem{matsyuk:1981}
    {\it Мацюк\ms Р.\ms Я.,} В~кн.: Граничные задачи математической физики.~Киев, Наукова думка, 1981. 79--81.
\bibitem{matsyuk:1982}{\it Мацюк\ms Р.\ms Я.,} В~зб.: Мат. методы и физ.-мех. поля. 1982. Вып.\ms 16. 84--88.
\bibitem{matsyuk:Dixon}
        {\it Dixon\ms W.\ms G.,}
        Proc. Roy. Soc. London. Ser.\ms A. 1970. {\bf 314}. 499--527.
\bibitem{NTSH2006}
    {\it Мацюк\ms Р.\ms Я.,} Фізичний збірник НТШ. 2006. т.\ms6. 206--114.
\end{thebibliography}
\end{document}